\pdfoutput=1
\newif\iffull\fulltrue
\documentclass{llncs}
\usepackage{llncsdoc}

\usepackage{caption}
\usepackage{prooftree}
\usepackage{soul}
\usepackage{tikz}
\usepackage{siunitx}
\usepackage{listings}
\usepackage{pifont}
\usepackage{multirow}
\usepackage{amsmath}
\usepackage{amsfonts}
\usepackage{amssymb}
\usepackage{graphicx}
\usepackage{paralist}
\usepackage{wrapfig}
\usepackage{balance}
\usepackage{dsfont}
\usepackage{enumitem}
\usepackage{mathtools}
\usepackage{listings}
\usepackage[plain]{algorithm}
\usepackage[noend]{algpseudocode}
\usepackage{pifont}
\usepackage{stmaryrd}
\usepackage{blindtext}
\usepackage{enumitem}
\usepackage{comment}
\usepackage{url}
\usepackage{etoolbox}
\usepackage{pbox}
\usepackage{csquotes}
\usepackage{ctable} 
\usepackage{textcase}
\usepackage{wasysym}
\usepackage{mdframed}

\usepackage[theorems,skins]{tcolorbox}
\usepackage{empheq}

\newtcolorbox{mymathbox}[1][]{colback=white, sharp corners, #1}

\newtcbox{\othermathbox}[1][]{nobeforeafter, math upper, tcbox raise base, enhanced,  colback=black!0, colframe=black!20,  left=1em, top=0em, right=3em, bottom=0em}

\renewcommand{\paragraph}[1]{\vspace{.05in}\noindent\textbf{{#1}~~}}

\usepackage{enumitem}

\makeatletter
\lst@InstallKeywords k{attributes}{attributestyle}\slshape{attributestyle}{}ld
\makeatother

\usepackage[scaled=0.80]{DejaVuSansMono}
\lstdefinestyle{customc}{
  belowcaptionskip=1\baselineskip,
  breaklines=true,
  xleftmargin=\parindent,
  language=C,
  showstringspaces=false,
  basicstyle=\ttfamily,
  keywordstyle=\bfseries\ttfamily\color{keywordcolor},
  commentstyle=\itshape\color{black},
  identifierstyle=\ttfamily\color{black},
  stringstyle=\itshape\color{NavyBlue},
  keywords={ map, flatMap, reduce, then, in, if, else, reduceByKey},
moreattributes={let, where}, 
attributestyle = \bfseries\ttfamily\color{attributecolor}
}

\setlist[description]{
   labelindent=.3cm,
   style=unboxed,
   leftmargin=.3cm,
   topsep=2pt,
   itemsep=1ex
}

 \definecolor{mypink3}{cmyk}{0, 0.7808, 0.4429, 0.1412}

\everypar{\looseness=-1}

\definecolor{lightgray}{gray}{0.9}
\definecolor{midgray}{gray}{0.65}
\definecolor{darkgray}{gray}{0.4}

\newcommand{\abr}[1]{\textsc{\MakeLowercase{#1}}}

\renewcommand{\vec}[1]{\boldsymbol{#1}}

\renewcommand{\leq}{\leqslant}
\renewcommand{\geq}{\geqslant}

\newcommand{\rone}{(\emph{i})~}
\newcommand{\rtwo}{(\emph{ii})~}
\newcommand{\rthree}{(\emph{iii})~}
\newcommand{\rfour}{(\emph{iv})~}

\definecolor{keywordcolor}{gray}{0.0}
\definecolor{attributecolor}{gray}{0.0}
\definecolor{wildcolor}{gray}{0.8}

\usepackage{relsize}

\newcommand\earr\hookrightarrow

\newcommand{\aset}{B}
\newcommand{\aelem}{b}
\newcommand{\avec}{\vec{b}}

\newcommand{\dist}{\mathit{dist}}

\newcommand{\supp}{\mathit{supp}}

\newcommand{\aspace}[1]{\leftrightsquigarrow^{#1}}

\newcommand{\prog}{P}
\newcommand{\vars}{V}
\newcommand{\states}{S}
\newcommand{\varsi}{\vars^I}

\newcommand{\expr}{\mathit{exp}}
\newcommand{\bexpr}{\mathit{bexp}}
\newcommand{\dexpr}{\mathit{dexp}}

\newcommand{\sem}[1]{\llbracket #1\rrbracket}

\newcommand{\post}{\mathsf{post}}
\newcommand{\cpost}{\mathsf{cpost}}

\newcommand{\eps}\epsilon

\newcommand{\enc}{\mathsf{enc}}

\algrenewcommand\algorithmicfunction{\textsf{\textbf{fun}}}

\newcommand{\true}{\mathit{true}}
\newcommand{\false}{\mathit{false}}

\renewcommand{\epsilon}{\varepsilon}

\newcommand{\bern}{\mathsf{bern}}

\begin{document}

\title{Constraint-Based Synthesis of Coupling Proofs\iffull\else\thanks{%
The full version of this paper is available at \protect\url{XYZ}.}\fi}

\author{Aws Albarghouthi\inst{1} \and Justin Hsu\inst{2,3}}
\institute{University of Wisconsin--Madison, Madison, WI
\and University College London, London, UK
\and Cornell University, Ithaca, NY}
\authorrunning{Aws Albarghouthi and Justin Hsu} 

\maketitle
\begin{abstract}
  \emph{Proof by coupling} is a classical technique for proving properties about
  pairs of randomized algorithms by carefully \emph{relating} (or
  \emph{coupling}) two probabilistic executions.  In this paper, we show how to
  automatically construct such proofs for probabilistic programs.  First, we
  present $f$-\emph{coupled postconditions}, an abstraction describing two
  correlated program executions.  Second, we show how properties of $f$-coupled
  postconditions can imply various probabilistic properties of the original
  programs.  Third, we demonstrate how to reduce the proof-search problem to a
  purely logical \emph{synthesis problem} of the form $\exists f \ldotp \forall
  X \ldotp \varphi$, making probabilistic reasoning unnecessary.  We develop a
  prototype implementation to automatically build coupling proofs for
  probabilistic properties, including uniformity and independence of program
  expressions.
\end{abstract}

\section{Introduction}\label{sec:intro}

In this paper, we aim to automatically synthesize \emph{coupling proofs} for
probabilistic programs and properties. Originally designed for proving
properties comparing two probabilistic programs---so-called \emph{relational
properties}---a coupling proof describes how to correlate two executions of the
given programs, simulating both programs with a single probabilistic program. By
reasoning about this combined, \emph{coupled} process, we can often give simpler
proofs of probabilistic properties for the original pair of programs.

A number of recent works have leveraged this idea to verify relational
properties of randomized algorithms, including differential
privacy~\cite{BKOZ13-toplas,BGGHS16,BGGHS16c}, security of cryptographic
protocols~\cite{DBLP:conf/popl/BartheFGSSB14}, convergence of Markov
chains~\cite{barthe2017coupling}, robustness of machine learning
algorithms~\cite{BartheEGHS18}, and more. Recently, Barthe et
al.~\cite{barthe2017proving} showed how to reduce certain \emph{non-relational}
properties---which describe a single probabilistic program---to relational
properties of two programs, by duplicating the original program or by
sequentially composing it with itself.

While coupling proofs can simplify reasoning about probabilistic properties,
they are not so easy to use; most existing proofs are carried out manually in
relational program logics using interactive theorem provers. In a nutshell, the
main challenge in a coupling proof is to select a correlation for each pair of
corresponding sampling instructions, aiming to induce a particular relation
between the outputs of the coupled process; this relation then implies the
desired relational property. Just like finding inductive invariants in proofs
for deterministic programs, picking suitable couplings in proofs can require
substantial ingenuity.

To ease this task, we recently showed how to cast the search for coupling proofs
as a program synthesis problem~\cite{DBLP:journals/pacmpl/AlbarghouthiH18},
giving a way to automatically find sophisticated proofs of differential privacy
previously beyond the reach of automated verification. In the present paper, we
build on this idea and present a general technique for constructing coupling
proofs, targeting \emph{uniformity} and \emph{probabilistic independence}
properties. Both are fundamental properties in the analysis of randomized
algorithms, either in their own right or as prerequisites to proving more
sophisticated guarantees; uniformity states that a randomized expression takes
on all values in a finite range with equal probability, while probabilistic
independence states that two probabilistic expressions are somehow
uncorrelated---learning the value of one reveals no additional information about
the value of the other.


Our techniques are inspired by the automated proofs of differential privacy we
considered previously~\cite{DBLP:journals/pacmpl/AlbarghouthiH18}, but the
present setting raises new technical challenges.
\begin{description}
  \item[Non-lockstep execution.]
    To prove differential privacy, the behavior of a single program is compared
    on two related inputs. To take advantage of the identical program structure,
    previous work restricted attention to \emph{synchronizing} proofs, where the
    two executions can be analyzed assuming they follow the same control flow
    path. In contrast, coupling proofs for uniformity and independence often
    require relating two programs with different shapes, possibly following
    completely different control flows~\cite{barthe2017proving}.

    To overcome this challenge, we take a different approach.
    Instead of incrementally finding couplings for corresponding
    pairs of sampling instructions---requiring the executions to be tightly
    synchronized---we first lift all sampling instructions to the front of the
    program and pick a coupling once and for all. The remaining execution
    of both programs can then be encoded separately, with no need for
    lockstep synchronization (at least for loop-free programs---looping programs
    require a more careful treatment).
  \item[Richer space of couplings.]
    The heart of a coupling proof is selecting---among multiple possible
    options---a particular correlation for each pair of random sampling
    instructions. Random sampling in differentially private programs typically
    use highly domain-specific distributions, like the Laplace distribution,
    which support a small number of useful couplings. Our prior work leveraged
    this feature to encode a collection of primitive couplings into the
    synthesis system. However, this is no longer possible when programs sample
    from distributions supporting richer couplings, like the uniform
    distribution. Since our approach coalesces all sampling instructions at the
    beginning of the program (more generally, at the head of the loop), we also
    need to find couplings for products of distributions.

    We address this problem in two ways. First, we allow couplings of two
    sampling instructions to be specified by an injective function $f$ from one
    range to another. Then, we impose requirements---encoded as standard logical
    constraints---to ensure that $f$ indeed represents a coupling; we call such
    couplings $f$-\emph{couplings}.
  \item[More general class of properties.]
    Finally, we consider a broad class of properties rather than just
    differential privacy. While we focus on uniformity and independence for
    concreteness, our approach can establish general equalities between products
    of probabilities, i.e., probabilistic properties of the form
    \[
      \prod_{i = 1}^m\Pr[ e_i \in E_i ]
      =
      \prod_{j = 1}^n\Pr[ e_j' \in E_j' ] ,
    \]
    where $e_i$ and $e_j'$ are program expressions in the first and second
    programs respectively, and $E_i$ and $E_j'$ are predicates.  As an example,
    we automatically establish a key step in the proof of Bertrand's Ballot
    theorem~\cite{feller1}.
\end{description}

\paragraph{Paper Outline.}
After overviewing our technique on a motivating example
(Section~\ref{sec:overview}), we detail our main contributions.
\begin{itemize}
  \item \textbf{Proof technique:}
  We introduce $f$-\emph{coupled postconditions}, a form of postcondition for
  two probabilistic programs where random sampling instructions in the two
  programs are correlated by a function $f$. Using $f$-coupled postconditions,
  we present proof rules for establishing uniformity and independence of program
  variables, fundamental properties in the analysis of randomized algorithms
  (Section~\ref{sec:proofrule}).

  \item \textbf{Reduction to constraint-based synthesis:} We demonstrate how to
    automatically find coupling proofs by transforming our proof rules into
    logical constraints of the form $\exists f \ldotp \forall X \ldotp
    \varphi$---a synthesis problem.  A satisfiable constraint shows the
    existence of a function $f$---essentially, a compact encoding of a coupling
    proof---implying the target property (Section~\ref{sec:synthesis}).

  \item \textbf{Extension to looping programs:}
  We extend our technique to reason about loops, by requiring synchronization at
  the loop head and finding a coupled invariant (Section~\ref{sec:loops}).

  \item \textbf{Implementation and evaluation:}
  We implement our technique and evaluate it on several case studies,
  automatically constructing coupling proofs for interesting properties of a
  variety of algorithms (Section~\ref{sec:eval}).
\end{itemize}
We conclude by comparing our technique with related approaches (Section~\ref{sec:related}).

\section{Overview and Illustration}\label{sec:overview}

\subsection{Introducing $f$-Couplings}
\paragraph{A Simple Example.}
We begin by illustrating $f$-couplings over two identical Bernoulli
distributions, denoted by the following \emph{probability mass functions}:
$\mu_1(x) = \mu_2(x) = 0.5$ for all $x \in \mathds{B}$ (where $\mathds{B} =
\{\true,\false\})$. In other words, the distribution $\mu_i$ returns $\true$
with probability $0.5$, and $\false$ with probability $0.5$.

An $f$-\emph{coupling} for $\mu_1,\mu_2$ is a function $f:
\mathds{B}\to\mathds{B}$ from the domain of the first distribution
($\mathds{B}$) to the domain of the second (also $\mathds{B}$); $f$
should be injective and satisfy the \emph{monotonicity property}: $\mu_1(x) \leq \mu_2(f(x))$ for all $x \in
\mathds{B}$. In other words, $f$ relates each element $x \in \mathds{B}$ with
an element $f(x)$ that has an equal or larger probability in $\mu_2$.
For example, consider the function $f_\neg$ defined as
\[ f_\neg(x) = \neg x . \]
This function relates $\true$ in $\mu_1$
with $\false$ in $\mu_2$, and vice versa.
Observe that $\mu_1(x) \leq \mu_2(f_\neg(x))$ for all $x \in \mathds{B}$,
satisfying the definition of an $f_\neg$-coupling.
We write $\mu_1 \aspace{f_\neg} \mu_2$ when there is an $f_\neg$-coupling
for $\mu_1$ and $\mu_2$.

\paragraph{Using $f$-Couplings.}
An $f$-coupling can imply useful properties about the
distributions $\mu_1$ and $\mu_2$. For example, suppose we want
to prove that $\mu_1(\true) = \mu_2(\false)$.
The fact that there is an $f_\neg$-coupling of $\mu_1$ and $\mu_2$ immediately
implies the equality: by the monotonicity property,
\begin{align*}
   \mu_1(\true) \leq \mu_2(f_\neg(\true)) = \mu_2(\false)\\
   \mu_1(\false) \leq \mu_2(f_\neg(\false)) = \mu_2(\true)
\end{align*}
and therefore $\mu_1(\true) = \mu_2(\false)$.
More generally, it suffices to find an
$f$-coupling of $\mu_1$ and $\mu_2$ such that
\[ \underbrace{\{(x,f(x)) \mid x\in\mathds{B}\}}_{\Psi_f} \subseteq \{(z_1,z_2) \mid z_1 = \true \iff z_2 = \false\} , \]
where $\Psi_f$ is induced by $f$; in particular, the
$f_\neg$-coupling satisfies this property.

\subsection{Simulating a Fair Coin}\label{ssec:fair}

\begin{wrapfigure}{r}{3.5cm}
  \vspace{-.45in}
  \smaller
  \center
    \begin{algorithmic}
      \Function{\sf fairCoin}{$p \in (0,1)$}
      \State $x \gets \false$
      \State $y \gets \false$
      \While{$x = y$}
        \State $x \sim \bern(p)$
        \State $y \sim \bern(p)$
      \EndWhile
      \State \Return $x$
      \EndFunction
    \end{algorithmic}
  \caption{Simulating a fair coin using an unfair one}\label{fig:fair}
\end{wrapfigure}

Now, let's use $f$-couplings to prove more interesting properties.  Consider the
program \textsf{fairCoin} in Figure~\ref{fig:fair}; the program simulates a fair
coin by flipping a possibly biased coin that returns $\true$ with probability $p
\in (0,1)$, where $p$ is a program parameter.  Our goal is to prove that for any
$p$, the output of the program is a uniform distribution---it simulates a
fair coin.  We consider two separate copies of \textsf{fairCoin} generating
distributions $\mu_1$ and $\mu_2$ over the returned value $x$ for the same bias
$p$, and we construct a coupling showing $\mu_1(\true) = \mu_2(\false)$, that
is, heads and tails have equal probability.

\paragraph{Constructing $f$-Couplings.}
At first glance, it is unclear how to construct an $f$-coupling; unlike the
distributions in our simple example, we do not have a concrete description of
$\mu_1$ and $\mu_2$ as uniform distributions (indeed, this is what we are trying
to establish). The key insight is that we do not need to construct our coupling
in one shot. Instead, we can specify a coupling for the concrete, primitive
sampling instructions in the body of the loop---which we know sample from
$\bern(p)$---and then extend to a $f$-coupling for the whole loop and $\mu_1,
\mu_2$.

For each copy of \textsf{fairCoin}, we coalesce the two sampling statements inside the loop
into a single sampling statement from the product distribution:
\[ x, y \sim \bern(p) \times \bern(p) \]
We have two such joint distributions $\bern(p) \times \bern(p)$ to couple, one
from each copy of \textsf{fairCoin}.  We use the following function $f_\mathit{swap} : \mathds{B}^2 \to \mathds{B}^2$:
\[ f_{\mathit{swap}}(x,y) = (y,x) \]
which exchanges the values of $x$ and $y$. Since this is an injective function
satisfying the monotonicity property
\[ (\bern(p) \times \bern(p))(x, y)
\leq (\bern(p) \times \bern(p))(f_{\mathit{swap}}(x, y)) \]
for all $(x, y) \in \mathds{B} \times \mathds{B}$ and $p \in (0,1)$, we have an
$f_{\mathit{swap}}$-coupling for the two copies of $\bern(p) \times \bern(p)$.

\paragraph{Analyzing the Loop.}
To extend a $f_{body}$-coupling on loop bodies to the entire loop, it suffices
to check a synchronization condition: the coupling from $f_{body}$ must ensure
that the loop guards are equal so the two executions synchronize at the loop
head.  This holds in our case: every time the first program executes the
statement $x, y \sim \bern(p) \times \bern(p)$, we can think of $x,y$ as
non-deterministically set to some values $(a,b)$, and the corresponding
variables in the second program as set to $f_{\mathit{swap}}(a,b) = (b,a)$. The
loop guards in the two programs are equivalent under this choice, since $a = b$
is equivalent to $b = a$, hence we can analyze the loops in lockstep.  In
general, couplings enable us to relate samples from a pair of probabilistic
assignments as if they were selected non-deterministically, often avoiding
quantitative reasoning about probabilities.

Our constructed coupling for the loop guarantees that \rone both
programs exit the loop at the same time, and \rtwo when the two programs
exit the loop, $x$ takes opposite values in the two programs. In other words,
there is an $f_{\mathit{loop}}$-coupling of $\mu_1$ and $\mu_2$ for some
function $f_{\mathit{loop}}$ such that
\begin{align}
  \Psi_{f_{\mathit{loop}}} \subseteq \{(z_1,z_2) \mid z_1 = \true \iff z_2 = \false\} ,
  \label{form:ex_unif}
\end{align}
implying $\mu_1(\true) = \mu_2(\false)$. Since both distributions
are output distributions of $\textsf{fairCoin}$---hence $\mu_1 = \mu_2$---we
conclude that \textsf{fairCoin} simulates a fair coin.

Note that our approach does not need to construct $f_{loop}$ concretely---this
function may be highly complex. Instead, we only
need to show that $\Psi_{f_{\mathit{loop}}}$ (or some over-approximation) lies
inside the target relation in Formula~\ref{form:ex_unif}.

\paragraph{Achieving Automation.}
%
Observe that once we have fixed an $f_\mathit{body}$-coupling for the sampling
instructions inside the loop body, checking that the $f_\mathit{loop}$-coupling
satisfies the conditions for uniformity (Formula~\ref{form:ex_unif}) is
essentially a program verification problem.  Therefore, we can cast the problem
of constructing a coupling proof as a logical problem of the form
$\exists f \ldotp \forall X \ldotp \varphi$, where $f$ is the $f$-coupling we
need to discover and $\forall X \ldotp \varphi$ is a constraint ensuring that
\rone $f$ indeed represents an $f$-coupling, and \rtwo the $f$-coupling implies
uniformity. Thus, we can use established synthesis-verification techniques to
solve the resulting constraints (see,
e.g.,~\cite{DBLP:conf/asplos/Solar-LezamaTBSS06,alur2013syntax,Beyene14}).


\section{A Proof Rule for Coupling Proofs} \label{sec:proofrule}

In this section, we develop a technique for constructing couplings and formalize
proof rules for establishing uniformity and independence properties over program
variables.  We begin with background on probability distributions and couplings.

\subsection{Distributions and Couplings}

\paragraph{Distributions.}
%
A function $\mu : \aset \to [0,1]$ defines a
\emph{distribution} over a countable set $\aset$ if $\sum_{\aelem \in \aset}
\mu(\aelem) = 1$. We will often write $\mu(A)$ for a subset $A
\subseteq \aset$ to mean $\sum_{x \in A} \mu(x)$. We write $\dist(\aset)$ for the set of all distributions over
$\aset$.

We will need a few standard constructions on distributions.  First, the
\emph{support} of a distribution $\mu$ is defined as  $\supp(\mu) = \{\aelem
\in \aset \mid \mu(\aelem) > 0\}$.  Second, for a distribution on pairs $\mu
\in \dist(\aset_1 \times \aset_2)$, the first and second \emph{marginals} of
$\mu$, denoted $\pi_1(\mu)$ and $\pi_2(\mu)$ respectively, are distributions
over $\aset_1$ and $\aset_2$:
\begin{align*}
  \pi_1(\mu)(\aelem_1) \triangleq \sum_{\aelem_2 \in \aset_2}\mu(\aelem_1,\aelem_2)
  \hspace{1in}
  \pi_2(\mu)(\aelem_2) \triangleq \sum_{\aelem_1 \in \aset_1}\mu(\aelem_1,\aelem_2) .
\end{align*}

\paragraph{Couplings.}
Let $\Psi \subseteq \aset_1 \times \aset_2$ be a binary relation.  A
$\Psi$-\emph{coupling} for distributions $\mu_1$ and $\mu_2$ over $\aset_1$ and
$\aset_2$  is a distribution $\mu \in \dist(\aset_1 \times \aset_2)$ with \rone
$\pi_1(\mu) = \mu_1$ and $\pi_2(\mu) = \mu_2$; and \rtwo $\supp(\mu) \subseteq
\Psi$. We write $\mu_1 \aspace{\Psi} \mu_2$ when there exists a $\Psi$-coupling
between $\mu_1$ and $\mu_2$.

An important fact is that an injective function $f: \aset_1 \to \aset_2$ where
$\mu_1(\aelem) \leq \mu_2(f(\aelem))$ for all $\aelem \in \aset_1$ induces a
coupling between $\mu_1$ and $\mu_2$; this follows from a general theorem by
Strassen~\cite{strassen1965existence}, see also \cite{JHThesis}.  We
write  $\mu_1 \aspace{f} \mu_2$ for $\mu_1 \aspace{\Psi_f} \mu_2$, where $\Psi_f
= \{(\aelem_1, f(\aelem_1)) \mid \aelem_1 \in \aset_1\}$.
%
%
The existence of a coupling can imply various useful properties about the two
distributions. The following general fact will be the most important for our
purposes---couplings can prove equalities between probabilities.

\begin{proposition} \label{prop:couple-iff-eq}
  Let $E_1 \subseteq \aset_1$ and $E_2 \subseteq \aset_2$ be two events, and let
  $\Psi_= \triangleq \{(\aelem_1, \aelem_2) \mid \aelem_1 \in E_1 \iff \aelem_2
  \in E_2\}$. If $\mu_1 \aspace{\Psi_=} \mu_2$, then $\mu_1(E_1) =
  \mu_2(E_2)$.
\end{proposition}

\subsection{Program Model}
Our program model uses an imperative language with probabilistic assignments,
where we can draw a random value from primitive distributions. We consider the
easier case of loop-free programs first; we consider looping programs in
Section~\ref{sec:loops}.

\paragraph{Syntax.}
A (loop-free) program $P$ is defined using the following grammar:
\begin{align*}
    P \coloneqq&~~  V \gets \expr & \text{(assignment)}\\
    & \mid V \sim \dexpr & \text{(probabilistic assignment)}\\
    & \mid \texttt{if } \bexpr \texttt{ then } P  \texttt{ else } P & \text{(conditional)}\\
    & \mid P; P& \text{(sequential composition)}
\end{align*}
where $\vars$ is the set of variables that can appear
in $\prog$, $\expr$ is an expression over $\vars$,
and $\bexpr$ is a Boolean expression over $\vars$.
A probabilistic assignment  samples from
a probability distribution defined by expression $\dexpr$;
for instance, $\dexpr$
might be $\bern(p)$, the Bernoulli distribution
with probability $p$ of returning $\true$.
We use $\varsi \subseteq \vars$ to denote the set of input program variables,
which are never assigned to.
All other variables are assumed to be defined before use.

We make a few simplifying assumptions. First, distribution expressions only
mention input variables $\varsi$, e.g., in the example above, $\bern(p)$, we
have $p\in\varsi$. Also, all programs are in \emph{static single assignment}
(\abr{SSA}) form, where each variable is assigned to only once and are
well-typed. These assumptions are relatively minor; they can can be verified
using existing tools, or lifted entirely at the cost of slightly more complexity
in our encoding.

\paragraph{Semantics.}
A state $s$ of a program $\prog$ is a valuation of all of its variables,
represented as a map from variables to values, e.g., $s(x)$ is the value of
$x\in\vars$ in $s$. We extend this mapping to expressions: $s(\expr)$ is the
valuation of $\expr$ in $s$, and $s(\dexpr)$ is the probability distribution
defined by $\dexpr$ in $s$.

We use $S$ to denote the set of all possible program states.  As is
standard~\cite{Kozen81}, we can give a semantics of $\prog$ as a function
$\sem{\prog}: S \to \dist(S)$ from states to distributions over states.  For an
output distribution $\mu = \sem{\prog}(s)$, we will abuse notation and write,
e.g., $\mu(x = y)$ to denote the probability of the event that the program
returns a state $s$ where $s(x = y) = \true$.

\paragraph{Self-Composition.}
We will sometimes need to simulate two separate executions of a program with a
single probabilistic program. Given a program $\prog$, we use $\prog_i$
to denote a program identical to $\prog$ but with all variables \emph{tagged}
with the subscript $i$.  We can then define the \emph{self-composition}: given a
program $\prog$, the program $\prog_1; \prog_2$ first executes $\prog_1$, and
then executes the (separate) copy $\prog_2$.

\subsection{Coupled Postconditions}

We are now ready to present the $f$-\emph{coupled postcondition},
an operator for approximating the outputs of two coupled programs.

\paragraph{Strongest Postcondition.}
We begin by defining a standard strongest postcondition
operator over single programs, treating probabilistic assignments as no-ops.
Given a set of states $Q \subseteq \states$, we define $\post$
as follows:
\begin{align*}
  \post(v \gets \expr, Q)  &= \{s[v \mapsto s(\expr)] \mid s \in Q\}\\
  \post(v \sim \dexpr, Q) &= Q\\
  \post(\texttt{if } \bexpr \texttt{ then } P  \texttt{ else } P', Q)  &=
    \{s' \mid s \in Q, s' \in \post(P,s), s(\bexpr) = \emph{true}\}\\
    &\cup \{s' \mid s \in Q, s' \in \post(P',s), s(\bexpr) = \emph{false}\}\\
  \post(P;P',Q) &= \post(P',\post(P,Q))
\end{align*}
where $s[v \mapsto c]$ is state $s$ with variable $v$ mapped to the value $c$.

\paragraph{$f$-Coupled Postcondition.}
We rewrite programs so that all probabilistic assignments are combined into a
single probabilistic assignment to a vector of variables appearing at the
beginning of the program, i.e., an assignment of the form $\vec{v} \sim \dexpr$
in $P$ and $\vec{v}' \sim \dexpr'$ in $P'$, where $\vec{v},\vec{v}'$ are vectors
of variables.  For instance, we can combine $x \sim \bern(0.5);
y\sim\bern(0.5)$ into the single statement $x,y \sim
\bern(0.5)\times\bern(0.5)$.

Let $\aset, \aset'$ be the domains of $\vec{v}$ and $\vec{v}'$,
$f: \aset \to \aset'$ be a function, and $Q \subseteq S \times S'$ be a set of
pairs of input states, where
$S$ and $S'$ are the states of $\prog$ and $\prog'$, respectively. We
define the $f$-coupled postcondition operator $\cpost$ as
\begin{align}
  \cpost(P, P', Q, f) &=
    \{ (\post(P, s), \post(P',s')) \mid (s,s') \in Q'\}
    \notag \\
  \text{where}&~ Q' = \{(s[\vec{v} \mapsto \avec], s'[\vec{v}' \mapsto f(\avec)]) \mid (s,s') \in Q, \vec{b} \in B\} ,
  \notag \\
  \text{assuming that}&~ \quad \forall  (s,s') \in Q \ldotp s(\dexpr) \aspace{f} s'(\dexpr') .
  \label{form:cond}
\end{align}
The intuition is that the values drawn from sampling assignments in both
programs are coupled using the  function $f$.  Note that this operation
non-deterministically assigns $\vec{v}$ from $P$ with some values $\vec{b}$, and
$\vec{v}'$ with $f(\vec{b})$.  Then, the operation simulates the executions of
the two programs.  Formula~\ref{form:cond} states that there is an $f$-coupling
for every instantiation of the two distributions used in probabilistic
assignments in both programs.

\begin{example}\label{ex:fcouple}
  Consider the  simple program $P$ defined as
  $x \sim \bern(0.5); x = \neg x$ and let
  $f_\neg(x) = \neg x$.
  Then, $\cpost(P,P,Q,f_\neg)$ is $\{(s,s') \mid s(x) = \neg s'(x)\}$.
\end{example}

The main soundness theorem shows there is a probabilistic coupling of the output
distributions with support contained in the coupled postcondition (we defer all
proofs to \iffull Appendix~\ref{app:proofs}\else the full version of this
paper\fi).

\begin{theorem}\label{thm:cpost}
  Let programs $P$ and $P'$ be of the form $\vec{v} \sim \dexpr; P_D$ and
  $\vec{v}' \sim \dexpr'; P'_D$, for deterministic programs $P_D, P'_D$.
  Given a function $f : B \to B'$ satisfying Formula~\ref{form:cond}, for every
  $(s,s') \in S \times S'$ we have $\sem{P}(s) \aspace{\Psi} \sem{P'}(s')$, where
  $\Psi = \cpost(P, P', (s,s'), f)$.
\end{theorem}

\subsection{Proof Rules for Uniformity and Independence}

We are now ready to demonstrate how to establish uniformity and independence of
program variables using $f$-coupled postconditions.  We will continue to assume
that random sampling commands have been lifted to the front of each program, and
that $f$ satisfies Formula~\ref{form:cond}.

\paragraph{Uniformity.}
Consider a program $\prog$ and a variable $v^* \in \vars$ of finite, non-empty domain $\aset$.
Let $\mu = \sem{\prog}(s)$ for some state $s\in S$.
We say that variable $v^*$ is \emph{uniformly distributed}
in $\mu$ if $\mu(v^* = \aelem) = \frac{1}{|\aset|}$ for every $\aelem \in
\aset$.

The following theorem connects uniformity with $f$-coupled postconditions.
\begin{theorem}[Uniformity]
\label{thm:uniform}
Consider a program $P$ with $\vec{v} \sim \dexpr$ as its first statement and a
designated return variable $v^* \in \vars$ with domain $\aset$. Let $Q = \{(s,s)
\mid s \in S\}$ be the input relation. If we have
\[
  \exists f \ldotp \cpost(P,P,Q,f) \subseteq \{ (s, s') \in S \times S \mid s(v^*) = \aelem \iff s'(v^*) = \aelem'\}
\]
for all $\aelem,\aelem' \in \aset$, then for any input $s \in S$ the final value
of $v^*$ is uniformly distributed over $\aset$ in $\sem{P}(s)$.
\end{theorem}

The intuition is that in the two $f$-coupled copies of $P$, the first $v^*$ is
equal to $b$ exactly when the second $v^*$ is equal to $b'$. Hence, the
probability of returning $b$ in the first copy and $b'$ in the second copy are
the same. Repeating for every pair of values $b,b'$, we conclude that $v^*$ is
uniformly distributed.

\begin{example}\label{ex:unif}
Recall Example~\ref{ex:fcouple} and
let $b = \true$ and $b' = \false$.
We have $$\cpost(P,P,Q,f_\neg) \subseteq
\{ (s, s') \in S \times S \mid s(x) = b \iff s'(x) = b'\}.$$
This is sufficient to prove uniformity (the case with $b = b'$ is trivial).
\end{example}

\paragraph{Independence.}
We now present a proof rule for independence.  Consider a program $P$ and two
variables $v^*,w^* \in \vars$ with domains $\aset$ and $\aset'$, respectively.
Let $\mu = \sem{\prog}(s)$ for some state $s\in S$.  We say that $v^*,w^*$ are
\emph{probabilistically independent} in $\mu$ if $\mu(v^* = \aelem \land w^* =
\aelem') = \mu(v^* = \aelem) \cdot \mu(w^* = \aelem')$ for every $\aelem \in
\aset$ and $\aelem' \in \aset'$.

The following theorem connects independence with $f$-coupled postconditions.  We
will self-compose two tagged copies of $P$, called $P_1$ and $P_2$.

\begin{theorem}[Independence] \label{thm:independence}
Assume a program $P$ and define the relation
\[
  Q = \{(s, s_1\oplus s_2) \mid s \in S, s_i \in S_i, s(v) = s_i(v_i), \text{for all } v\in \varsi\} ,
\]
where $\oplus$ takes the union of two maps with disjoint domains.  Fix some
$w^*, v^* \in \vars$ with domains $\aset,\aset'$, and assume that for all
$\aelem \in \aset$, $\aelem' \in \aset'$, there exists a function $f$ such that
$\cpost(P,(P_1;P_2),Q,f)$ is contained in
\[
  \{ (s', s_1' \oplus s_2') \mid s'(v^*) = \aelem \land s'(w^*) = \aelem' \iff s_1'(v^*_1) = \aelem \land s_2'(w^*_2) = \aelem' \} .
\]
Then, $w^*,v^*$ are independently distributed in $\sem{P}(s)$ for all inputs $s
\in S$.\iffull\footnote{%
  Refer to Appendix~\ref{app:condindep} for a similar formulation of conditional independence.}\fi
\end{theorem}

The idea is that under the coupling, the probability of $P$ returning $v^* =
\aelem \land w^* = \aelem'$ is the same as the probability of $P_1$ returning
$v^* = b$ and $P_2$ returning $w^* = b'$, for all values $b,b'$.  Since $P_1$
and $P_2$ are two independent executions of $P$ by construction, this
establishes independence of $v^*$ and $w^*$.

\section{Constraint-Based Formulation of Proof Rules} \label{sec:synthesis}
In Section~\ref{sec:proofrule}, we formalized the problem of constructing a
coupling proof using $f$-coupled postconditions. We now automatically find such
proofs by posing the problem as a constraint, where a solution gives a function
$f$ establishing our desired property.

\subsection{Generating Logical and Probabilistic Constraints}

\paragraph{Logical Encoding.}
We first encode program executions as formulas in first-order logic, using the
following encoding function:
\begin{align*}
  \enc(v \gets \expr) &\triangleq v = \expr\\
  \enc(v \sim \dexpr) &\triangleq \true\\
  \enc(\texttt{if } \bexpr \texttt{ then } P  \texttt{ else } P')
  &\triangleq (\bexpr \Rightarrow \enc(P)) \land (\neg \bexpr \Rightarrow \enc(P'))\\
    \enc(P;P') &\triangleq \enc(P) \land \enc(P')
\end{align*}
We assume a direct correspondence between expressions in our language and the
first-order theory used for our encoding, e.g., linear arithmetic. Note that the
encoding disregards probabilistic assignments, encoding them as $\true$; this
mimics the semantics of our strongest postcondition operator $\post$.
Probabilistic assignments will be handled via a separate encoding of
$f$-couplings.

As expected, \textsf{enc} reflects the strongest postcondition \textsf{post}.

\begin{lemma} \label{lem:enc-sound}
  Let $P$ be a program and let $\rho$ be any assignment of the variables. An
  assignment $\rho'$ agreeing with $\rho$ on all input variables $V^I$ satisfies
  the constraint $\enc(P)[\rho'/V]$ precisely when $\post(P, \{\rho\}) = \{ \rho'
  \}$, treating $\rho,\rho'$ as program states.
\end{lemma}

\paragraph{Uniformity Constraints.}
We can encode the conditions in Theorem~\ref{thm:uniform} for showing uniformity
as a logical constraint. For a program $P$ and a copy $P_1$, with first
statements $\vec{v} \sim \dexpr$ and $\vec{v}_1 \sim \dexpr_1$, we define the
constraints:
\begin{empheq}[box=\othermathbox]{align}
  \forall a, a' \ldotp & \exists f \ldotp \forall \vars, \vars_1 \ldotp \nonumber \label{c:main} \\
  & (\varsi = \varsi_1 \land \vec{v}_1 = f(\vec{v}) \land \enc(P) \land \enc(P_1)) \\
  &~~~~~~~~~~~\Longrightarrow (v^* = a \iff v^*_1=a') \nonumber\\
  &\varsi = \varsi_1 \Longrightarrow \dexpr \aspace{f} \dexpr_1 \label{c:couple}
\end{empheq}
Note that this is a second-order formula, as it quantifies over the
\emph{uninterpreted function} $f$.  The left side of the implication in
Formula~\ref{c:main} encodes an $f$-coupled execution of $P$ and $P_1$, starting
from equal initial states.  The right side of this implication encodes
the conditions for uniformity, as in Theorem~\ref{thm:uniform}.

Formula~\ref{c:couple} ensures that there is an $f$-coupling between $\dexpr$
and $\dexpr_1$ for any initial state; recall that $\dexpr$ may mention
input variables $\varsi$. The constraint $\dexpr \aspace{f} \dexpr_1$ is not a
standard logical constraint---intuitively, it is satisfied if $\dexpr \aspace{f}
\dexpr_1$ holds for some interpretation of $f$, $\dexpr$, and
$\dexpr_1$.\iffull\footnote{%
  Refer to Appendix~\ref{app:sem} for more on the semantics of coupling constraints.}\fi

\begin{example}
  The constraint
  \[ \exists f \ldotp \forall p, p' \ldotp p = p' \Rightarrow \bern(p) \aspace{f} \bern(p') \]
  holds by setting $f$ to the identity function $\text{id}$, since for any $p =
  p'$ we have an $f$-coupling $\bern(p) \aspace{\text{id}} \bern(p')$.
\end{example}

\begin{example}
  Consider the program $x \sim \bern(0.5); y = \neg x$.
  The constraints for uniformity of $y$ are
  \begin{align*}
    \forall a, a' \ldotp  \exists f \ldotp \forall \vars, \vars_1 \ldotp &
     ( x_1 = f(x) \land y = \neg x \land y_1 = \neg x_1) \Longrightarrow (y = a \iff y_1=a') \\
    & \bern(0.5) \aspace{f} \bern(0.5) .
  \end{align*}
  Since there are no input variables, $\varsi = \varsi_1$ is equivalent to $\true$.
\end{example}
\begin{theorem}[Uniformity constraints] \label{thm:unif-sound}
  Fix a program $P$ and variable $v^* \in \vars$. Let $\varphi$ be the
  uniformity constraints in Formulas~\ref{c:main} and \ref{c:couple}.  If
  $\varphi$ is valid, then $v^*$ is uniformly distributed in $\sem{P}(s)$ for
  all $s \in \states$.
\end{theorem}

\paragraph{Independence Constraints.}
Similarly, we can characterize independence constraints using the conditions in
Theorem~\ref{thm:independence}. After transforming the program $P_1; P_2$ to
start with the single probabilistic assignment statement $\vec{v}_{1,2} \sim
\dexpr_{1,2}$, combining probabilistic assignments in $P_1$ and $P_2$, we define
the constraints:
\begin{empheq}[box=\othermathbox]{align}
  \forall a, a' \ldotp & \exists f \ldotp \forall \vars, \vars_1, \vars_2 \ldotp \nonumber\\
                       & (\varsi = \varsi_1 = \varsi_2 \land \vec{v}_{1,2} = f(\vec{v}) \land \enc(P) \land \enc(P_1;P_2)) \label{c:indep:main} \\
                       &~~~~~~~\Longrightarrow (v^* = a \land w^*=a' \iff v^*_1 = a \land w^*_2 = a')\nonumber\\
  &\varsi = \varsi_1 = \varsi_2 \Longrightarrow \dexpr \aspace{f} \dexpr_{1,2} \label{c:indep:couple}
\end{empheq}

\begin{theorem}[Independence constraints] \label{thm:indep-sound}
  Fix a program $P$ and two variables $v^*,w^* \in \vars$. Let $\varphi$ be the
  independence constraints from Formulas~\ref{c:indep:main}
  and~\ref{c:indep:couple}.  If $\varphi$ is valid, then $v^*,w^*$ are
  independent in $\sem{P}(s)$ for all $s \in \states$.
\end{theorem}

\subsection{Constraint Transformation}\label{sec:trans}
To solve our constraints, we transform our constraints into the form $\exists
f\ldotp \forall X \ldotp \varphi$, where $\varphi$ is a first-order formula.
Such formulas can be viewed as \emph{synthesis problems}, and are often solvable
automatically using standard techniques.

We perform our transformation in two steps. First, we transform our constraint
into the form $\exists f\ldotp \forall X \ldotp \varphi_p$, where $\varphi_p$
still contains the coupling constraint.  Then, we replace the coupling
constraint with a first-order formula by logically encoding primitive
distributions as uninterpreted functions.

\paragraph{Quantifier Reordering.}
Our constraints are of the form $\forall a, a' \ldotp \exists f \ldotp \forall X
\ldotp \varphi$. Intuitively, this means that for \emph{every} possible value of
$a,a'$, we want \emph{one} function $f$ satisfying $\forall X \ldotp \varphi$. 
We can pull the existential quantifier
$\exists f$ to the outermost level by extending the function with additional
parameters for $a,a'$, thus defining a different function for every
interpretation of $a,a'$.  For the uniformity constraints this transformation
yields the following formulas:
\begin{empheq}[box=\othermathbox]{align}
    \exists g \ldotp \forall a, a' & \ldotp \forall \vars, \vars_1 \ldotp \nonumber \label{c:main:trans} \\
  & (\varsi = \varsi_1 \land \vec{v}_1 = g(a,a',\vec{v}) \land \enc(P) \land \enc(P_1)) \\
  &~~~~~~~~~~~\Longrightarrow (v^* = a \iff v^*_1=a') \nonumber\\
  &\varsi = \varsi_1 \Longrightarrow \dexpr \aspace{g(a,a',-)} \dexpr_1 \label{c:couple:trans}
\end{empheq}
where $g(a,a',-)$ is the function after partially applying $g$.

\paragraph{Transforming Coupling Constraints.}
Our next step is to eliminate coupling constraints.  To do so, we use the
definition of $f$-coupling, which states that $\mu_1 \aspace{f} \mu_2$ if  \rone
$f$ is injective and \rtwo $\forall x \ldotp \mu_1(x) \leq \mu_2(f(x))$. The
first constraint (injectivity) is straightforward. For the second point
(monotonicity), we can encode distribution expressions---which represent
functions to reals---as uninterpreted functions, which we then further
constrain.  For instance, the coupling constraint $\bern(p) \aspace{f}
\bern(p')$ can be encoded as
\begin{align*}
\forall x,y & \ldotp x \neq y \Rightarrow f(x) \neq f(y) & (\text{injectivity})\\
\forall x& \ldotp h(x) \leq h'(f(x)) &(\text{monotonicity})\\
\forall x& \ldotp \mathit{ite}(x = \true, h(x) = p, h(x) = 1-p) &(\bern(p) \text{ encoding})\\
\forall x& \ldotp \mathit{ite}(x = \true, h'(x) = p', h'(x) = 1-p')&(\bern(p') \text{ encoding})
\end{align*}
where $h, h': \mathds{B} \to \mathds{R}^{\geq 0}$ are uninterpreted functions
representing the probability mass functions of $\bern(p)$ and $\bern(p')$; note
that the third constraint encodes the distribution $\bern(p)$, which returns
$\true$ with probability $p$ and false with probability $1-p$, and the fourth
constraint encodes $\bern(p')$.

Note that if we cannot encode the definition of the distribution in our
first-order theory (e.g., if it requires non-linear constraints), or if we do
not have a concrete description of the distribution, we can simply elide the
last two constraints and under-constrain $h$ and $h'$. In
Section~\ref{sec:eval} we use this feature to prove properties of a program
encoding a Bayesian network, where the primitive distributions are unknown
program parameters.

\begin{theorem}[Transformation soundness]\label{thm:trans}
  Let $\varphi$ be the constraints generated for some program $P$.
  Let $\varphi'$ be the result of applying the above transformations
  to $\varphi$. If $\varphi'$ is valid, then $\varphi$ is valid.
\end{theorem}

\paragraph{Constraint Solving.}
After performing these transformations, we finally arrive at
constraints of the form $\exists g \ldotp \forall a,a' \ldotp \forall V
\ldotp \varphi$, where $\varphi$ is a first-order formula. These exactly match
constraint-based program synthesis problems. In Section~\ref{sec:eval}, we
use \abr{SMT} solvers and enumerative synthesis to handle these constraints.

\section{Dealing with Loops}\label{sec:loops}
So far, we have only considered loop-free programs.
In this section, we our approach to programs with loops.

\newcommand{\plf}{\prog^{\emph{b}}}
\paragraph{$f$-Coupled Postconditions and Loops.}
We consider programs of the form
\[
\textsf{while } \bexpr\ \plf
\]
where $\plf$ is a loop-free program that begins with the statement $\vec{v}\sim
\dexpr$; our technique can also be extended to handle nested loops. We assume
all programs terminate with probability 1 for any initial state; there are
numerous systems for verifying this basic property automatically (see, e.g.,
\cite{Chatterjee2016,Chatterjee:2016:AAQ:2837614.2837639,Chatterjee:2017:SIP:3009837.3009873}).
To extend our $f$-coupled postconditions, we let $\cpost(P, P', Q, f)$ be the
smallest set $I$ satisfying:
\begin{align}
  &Q \subseteq I \tag{initiation} \\
  &\cpost(\plf, {\plf}', I_\emph{en}, f) \subseteq I \tag{consecution} \\
  &I \subseteq \{s(\bexpr) = s'(\bexpr') \mid s \in S, s' \in S'\} \tag{synchronization}
\end{align}
where $I_{\emph{en}} \triangleq \{ (s,s') \in I \mid s(\bexpr) = \true\}$.

Intuitively, the set $I$ is the least inductive invariant for the two coupled
programs running with synchronized loops. Theorem~\ref{thm:cpost}, which
establishes that $f$-coupled postconditions result in couplings over output
distributions, naturally extends to a setting with loops.

\paragraph{Constraint Generation.}
To prove uniformity, we generate constraints much like the loop-free case except
that we capture the invariant $I$, modeled as a relation over the variables of
both programs, using a \emph{Constrained Horn-Clause} (\abr{CHC}) encoding.  As
is standard, we use $V', V_1'$ to denote primed copies of program variables
denoting their value after executing the body, and we assume that $\enc(P^b)$
encodes a loop-free program as a transition relation from states over $V$ to
states over $V'$.
\begin{footnotesize}
\begin{align*}
  \forall a, a' \ldotp & \exists f, I \ldotp \forall V, V_1, V', V_1' \ldotp & \\
  &  \varsi=\varsi_1 \Longrightarrow I(\vars,\vars_1) & \text{(initiation)}\\
  &   I(\vars,\vars_1) \land \bexpr \land
   \vec{v}_1' = f(\vec{v}') \land \enc(\plf) \land \enc(\plf_1) \Longrightarrow  I(\vars',\vars_1') & \text{(consecution)}\\
   & I(\vars,\vars_1) \Longrightarrow \bexpr = \bexpr_1 & \text{(synchronization)}\\
  & I(V,V_1) \Longrightarrow \dexpr \aspace{f} \dexpr_1 & \text{(coupling)}\\
  & I(V,V_1) \land \neg \bexpr \Longrightarrow (v^* = a \iff v^*_1 = a') & \text{(uniformity)}
\end{align*}
\end{footnotesize}

The first three constraints encode the definition of $\cpost$;
 the last two ensure that $f$ constructs a coupling and
that the invariant implies the uniformity condition when the loop terminates.
Using the technique presented in Section~\ref{sec:trans},
we can transform these constraints into the form $\exists f, I \ldotp \forall X \ldotp \varphi$.
That is, in addition to discovering the function $f$, we need to discover the invariant $I$.
\iffull
Appendix~\ref{app:encex} details the encoding of the fair coin example from
Section~\ref{ssec:fair}.
\fi

Proving  independence in looping programs poses additional
challenges, as directly applying the self-composition construction from
Section~\ref{sec:proofrule} requires relating a single loop with two loops. When the
number of loop iterations is deterministic, however, we may simulate two
sequentially composed loops with a single loop that interleaves the iterations
(known as \emph{synchronized} or \emph{cross}
product~\cite{ZaksP08,BartheCK11}) so that we reduce the synthesis problem to
finding a coupling for two loops.


\section{Implementation and Evaluation} \label{sec:eval}
We now discuss our implementation and five case studies
used for evaluation.

\begin{figure}[t!]
  \smaller
  \centering
  \begin{minipage}[t]{.3\textwidth}
    \begin{algorithmic}
      \Function{\sf fairCoin}{$p \in (0,1)$}
      \State $x \gets \false$
      \State $y \gets \false$
      \While{$x = y$}
        \State $x \sim  \bern(p)$
        \State $y \sim \bern(p)$
      \EndWhile
      \State \Return $x$
      \EndFunction
      \State
      \Function{\sf fairDie}{}
        \State{$x \gets \false$}
        \State{$y \gets \false$}
        \State{$z \gets \false$}
        \While{$x = y = z$}
          \State{$x \sim \bern(0.5)$}
          \State{$y \sim \bern(0.5)$}
          \State{$z \sim \bern(0.5)$}
        \EndWhile
        \Return{$(x, y, z)$}
      \EndFunction
    \end{algorithmic}
  \end{minipage}%
  \vrule
  \begin{minipage}[t]{.34\textwidth}
    \begin{algorithmic}
      \Function{\sf noisySum}{$n, p \in (0,1)$}
      \State{$\mathit{sum} \gets 0$}
        \For{$i = 1, \dots, n$}
          \State{$\mathit{noise}[i] \sim \bern(p)$}
          \State{$\mathit{sum} \gets \mathit{sum} + \mathit{noise}[i]$}
        \EndFor
        \State\Return{$\mathit{sum}$}
      \EndFunction
      \State\State
      \Function{\sf bayes}{$\mu, \mu', \mu''$}
        \State{$x \sim \mu$}
        \State{$y \sim \mu'$}
        \State{$z \sim \mu''$}
        \State{$w \gets f(x, y)$}
        \State{$w' \gets g(y, z)$}
        \State\Return{$(w, w')$}
      \EndFunction
    \end{algorithmic}
  \end{minipage}
  \vrule
  \begin{minipage}[t]{.35\textwidth}
    \begin{algorithmic}
      \State
      \Function{\sf ballot}{$n$}
      \State{$\emph{tie} \gets \false$}
      \State{$x_A \gets 0$}
      \State{$x_B \gets 0$}
      \For{$i = 1, \dots, n$}
        \State{$r \sim \bern(0.5)$}
        \If{$r = 0$}
          \State{$x_A \gets x_A + 1$}
        \Else
          \State{$x_B \gets x_B + 1$}
        \EndIf
        \If{$i = 1$}
          \State{$\mathit{first} \gets r$}
        \EndIf
        \If{$x_A = x_B$}
          \State{$\mathit{tie} \gets \true$}
        \EndIf
      \EndFor
      \State\Return{$(\mathit{first}, \mathit{tie})$}
      \EndFunction
    \end{algorithmic}
  \end{minipage}%
  \caption{Case study programs}
  \vspace{-.1in}
  \label{fig:algs}
\end{figure}

\paragraph{Implementation.}
To solve formulas of the form $\exists f \ldotp \forall X \ldotp \varphi$,
we implemented a simple solver using a \emph{guess-and-check} loop:
We iterate through various interpretations of $f$,
insert them into the formula, and check whether the resulting formula is valid.
In the simplest case,
we are searching for a function $f$ from $n$-tuples to $n$-tuples.
For instance, in Section~\ref{ssec:fair}, we discovered
the function $f(x,y) = (y,x)$.
Our implementation is parameterized by a grammar defining an infinite
set of interpretations of $f$, which involves permuting the arguments (as
above), conditionals, and other basic operations (e.g., negation for Boolean
variables).\iffull\footnote{%
We detail this grammar in Appendix~\ref{app:grammar}.}\fi{}
For checking validity of $\forall X \ldotp \varphi$ given $f$,
we use the Z3 \abr{SMT} solver~\cite{de2008z3} for loop-free programs.
For loops, we use an existing constrained-Horn-clause solver based
on the MathSAT \abr{SMT} solver~\cite{mathsat}.

\paragraph{Benchmarks and Results.}
As a set of case studies for our approach, we use 5 different programs collected
from the literature and presented in Figure~\ref{fig:algs}.  For these programs,
we prove uniformity, (conditional) independence properties, and other
probabilistic equalities.  For instance, we use our implementation to prove a
main lemma for the Ballot theorem~\cite{feller1}, encoded as the program
\textsf{ballot}.

Figure~\ref{fig:results} shows the time and number of loop iterations required by our implementation to discover a coupling proof.
The small number of iterations and time needed demonstrates the simplicity
of the discovered proofs.
For instance, the \textsf{ballot} theorem was proved in 3 seconds and only 4 iterations, while the \textsf{fairCoin} example (illustrated in  Section~\ref{ssec:fair})
required only two iterations and 1.4 seconds.
In all cases, the size of the synthesize function $f$ in terms of depth of its \abr{AST}
is no more than 4.
We describe these programs and properties in a bit more detail.

\paragraph{Case Studies: Uniformity (\textsf{fairCoin}, \textsf{fairDie}).}
The first two programs produce uniformly random values.  Our
approach synthesizes a coupling proof certifying uniformity for both of these
programs. The first program
\textsf{fairCoin}, which we saw in Section~\ref{ssec:fair}, produces a fair coin
flip given access to biased coin flips by repeatedly flipping two coins while
they are equal, and returning the result of the first coin as soon as the flips
differ. Note that the bias of the coin flip is a program parameter, and not
fixed statically.  The synthesized coupling swaps the result of the two samples,
mapping the values of $(x, y)$ to $(y, x)$.

\begin{wrapfigure}{r}{3.5cm}
  \vspace{-.1in}
  \smaller
  \centering
    \begin{tabular}{lcc}
      \toprule
      Program & Iters. & Time(s) \\
      \midrule
      \textsf{fairCoin} & 2 & 1.4 \\
      \textsf{fairDie} & 9 & 6.1 \\
      \textsf{noisySum} & 4 & 0.2 \\
      \textsf{bayes} & 5 & 0.4 \\
      \textsf{ballot} & 4 & 3.0 \\
      \bottomrule
    \end{tabular}
  \caption{Statistics} \label{fig:results}
\end{wrapfigure}

The second program \textsf{fairDie} gives a different construction for
simulating a roll of a fair die given fair coin flips. Three fair coins are
repeatedly flipped as long as they are all equal; the returned triple is the
binary representation of a
number in $\{ 1, \dots, 6 \}$, the result of the simulated roll. The
synthesized coupling is a bijection on triples of booleans $\mathds{B} \times
\mathds{B} \times \mathds{B}$; fixing any two possible output triples $(b_1,
b_2, b_3)$ and $(b_1', b_2', b_3')$ of distinct booleans, the coupling maps
$(b_1, b_2, b_3) \mapsto (b_1', b_2', b_3')$ and vice versa, leaving all other
triples unchanged.

\paragraph{Case Studies: Independence (\textsf{noisySum}, \textsf{bayes}).}
In the next two programs, our approach synthesizes coupling proofs of
independence and conditional independence of program variables in the output
distribution. The first program, \textsf{noisySum}, is a stylized program
inspired from privacy-preserving algorithms that sum a series of noisy samples;
for giving accuracy guarantees, it is often important to show that the noisy
draws are probabilistically independent.  We show that any pair of samples are
independent.

The second program, \textsf{bayes}, models a simple Bayesian network with three
independent variables $x, y, z$ and two dependent variables $w$ and $w'$,
computed from $(x, y)$ and $(y, z)$ respectively. We want to show that $w$ and
$w'$ are independent conditioned on any value of $y$; intuitively, $w$ and $w'$
only depend on each other through the value of $y$, and are independent
otherwise. We use a constraint encoding similar to the encoding for showing
independence \iffull (detailed in Appendix~\ref{app:condindep}) \fi to find a
coupling proof of this fact. Note that the distributions $\mu, \mu', \mu''$ of
$x, y, z$ are unknown parameters, and the functions $f$ and $g$ are also
uninterpreted. This illustrates the advantage of using a constraint-based
technique---we can encode unknown distributions and operations as uninterpreted
functions.

\paragraph{Case Studies: Probabilistic Equalities (\textsf{ballot}).}
As we mentioned in Section~\ref{sec:intro}, our approach extends naturally to proving
general probabilistic equalities beyond uniformity and independence. To
illustrate, we consider a lemma used to prove Bertrand's Ballot
theorem~\cite{feller1}. Roughly speaking, this theorem considers counting
ballots one-by-one in an election where there are $n_A$ votes cast for candidate
$A$ and $n_B$ votes cast for candidate $B$, where $n_A, n_B$ are parameters. If
$n_A > n_B$ (so $A$ is the winner) and votes are counted in a uniformly random
order, the Ballot theorem states that the probability that $A$ leads throughout
the whole counting process---without any ties---is precisely $(n_A - n_B) /
(n_A + n_B)$.

One way of proving this theorem, sometimes called Andr\'e's reflection
principle, is to show that the probability of counting the first vote for $A$ and
reaching a tie is equal to the probability of counting the first vote for $B$
and reaching a tie. We simulate the counting process slightly
differently---instead of drawing a uniform order to count the votes, our program
draws uniform samples for votes---but the original target property is equivalent
to the equality
\begin{equation} \label{eq:andre}
  \Pr[ \mathit{first}_1 = 0 \land \mathit{tie}_1 \land \psi(x_{A1}, x_{B1}) ]
  = \Pr[ \mathit{first}_2 = 1 \land \mathit{tie}_2 \land \psi(x_{A2}, x_{B2}) ]
\end{equation}
with $\psi(x_{Ai}, x_{Bi})$ is $x_{Ai} = n_A \land x_{Bi} = n_B$. Our approach
synthesizes a coupling and loop invariant showing that the coupled
post-condition is contained in
\[
  \{
    (s_1, s_2) \mid
    s_1(\mathit{first} = 0 \land \mathit{tie} \land \psi(x_A, x_B))
    \iff
    s_2(\mathit{first} = 0 \land \mathit{tie} \land \psi(x_A, x_B))
  \} ,
\]
giving Formula~\eqref{eq:andre} by Proposition~\ref{prop:couple-iff-eq} (see
Barthe et al.~\cite{barthe2017proving} for more details).

\section{Related Work}\label{sec:related}

Probabilistic programs have been a long-standing target of formal verification. 
We compare
with two of the most well-developed lines of research: probabilistic model
checking and deductive verification via program logics or expectations.

\paragraph{Probabilistic Model Checking.}
Model checking has proven to be a powerful tool for verifying probabilistic
programs, capable of automated proofs for various probabilistic properties
(typically encoded in probabilistic temporal logics);
there are now numerous mature implementations (see, e.g.,
\cite{forejt2011automated} or \cite[Ch.~10]{baier2008principles} for more
details). In comparison, our approach has the advantage of being fully
constraint-based. This gives it a number of unique features:
\rone it applies to programs with unknown inputs and variables over infinite domains;
\rtwo it applies to programs sampling from distributions with parameters, or
even ones sampling from unknown distributions modeled as uninterpreted functions
in first-order logic;
\rthree it applies to distributions over infinite domains; and
\rfour the generated coupling proofs are compact.
At the same time, our approach is specialized to the coupling proof technique
and is likely to be more incomplete.

\paragraph{Deductive Verification.}
Compared to general deductive verification systems for probabilistic programs,
like program logics~\cite{RandZ15,BEGGHS16,Hartog:thesis,Chadha07} or techniques
reasoning by pre-expectations~\cite{Morgan:1996}, the main benefit of our
technique is automation---deductive verification typically requires an
interactive theorem prover to manipulate complex probabilistic invariants. In
general, the coupling proof method limits reasoning about probabilities and
distributions to just the random sampling commands; in the rest of the program,
the proof can avoid quantitative reasoning entirely. As a result, our system can
work with non-probabilistic invariants and achieve full automation. Our approach
also smoothly handles properties involving the probabilities of multiple events,
like probabilistic independence, unlike techniques that analyze probabilistic
events one-by-one.

\paragraph{Acknowledgements.}
We thank Samuel Drews, Calvin Smith, and the anonymous reviewers for their helpful comments. Justin Hsu was
partially supported by ERC grant \#679127 and NSF grant \#1637532.
Aws Albarghouthi was supported by NSF grants \#1566015, \#1704117, and \#1652140.

\bibliographystyle{splncs03}
\bibliography{header,refs}

\iffull
\appendix
\section{Proof Rule for Conditional Independence}\label{app:condindep}
We formalize the proof rule for conditional independence
as follows.
Note that for conditional independence,
we need to coupled two self-composed copies of a program $P$.

\begin{theorem}[Conditional independence] \label{thm:condindep}
  Assume a program $\prog$ and a set
  \[
  Q = \{(s_1 \oplus s_2, s_3 \oplus s_4) \mid s_i(v_i) = s_j(v_j), \text{ for all } v \in \varsi\}
  \]
  Fix some $v^*, w^*, x^* \in \vars$ over domain $B$.
  Assume that for all $a,b,c \in B$,
  \begin{align*}
  \exists f \ldotp \cpost((\prog_1;\prog_2), (\prog_3;\prog_4), Q, f)
  \mathrel{\subseteq}
  \{(s_1 \oplus s_2, s_3 \oplus s_4) \mid \phi \}
\end{align*}
where
\begin{align*}
  \phi &\triangleq
  s_1(v^*) = a \land s_1(w^*) = b \land s_1(x^*) = s_2(x^*) = c\\
  &\iff s_3(v^*) = a \land s_4(w^*) = b \land s_3(x^*) = s_4(x^*) = c .
\end{align*}
Then for all $s \in S$, $v^*$ and $w^*$ are independent in
$\sem{\prog}(s)$, conditioned on $x^*$.
\end{theorem}

\section{Semantics of Coupling Constraints}\label{app:sem}
A coupling constraint is of the form
$\dexpr \aspace{f} \dexpr'$,
where $\dexpr, \dexpr'$
are two distribution expressions,
which are functions from some domain to distributions.
For instance, $\bern(x)$
is a function in $[0,1] \to \dist(\mathds{B})$;
i.e., given a real value in [0,1],
it returns a distribution over Booleans.
We assume all distribution expressions are parameterized
by variables mapping to interpreted first-order theories,
e.g., integers and reals.

A interpretation $m$ of the coupling constraint maps
variables in $\dexpr$ to values and gives an interpretation
to the uninterpreted function $f$.
We say that $m$ is a model of the coupling constraint
if $m(\dexpr) \aspace{m(f)} m(\dexpr')$ holds,
where $m(\dexpr)$ is the distribution expression instantiated
with the values of variables given in $m$.

\begin{example}
  Consider
  $\bern(x) \aspace{f} \bern(x)$.
  Let $m$ be interpretation that sets $x$ to 0.5
  and $f$ to the function s.t. $\forall y \ldotp f(y) = \neg y$.
  Then, $m$ is a model of the coupling constraint.
\end{example}

\section{Proofs}
\label{app:proofs}

\paragraph{Proof of Theorem~\ref{thm:cpost}.}
  Supposing that $P = \vec{v} \sim \dexpr; P_D$ and $P' = \vec{v}' \sim \dexpr';
  P_D'$, we have $\post(P, s) = \post(P_D, s)$ and $\post(P', s') = \post(P_D',
  s')$, and so
  \begin{align*}
    \cpost(P, P', (s, s'), f) &=
    \{ (\post(P_D, s), \post(P'_D,s')) \mid (s,s') \in Q'\}   \\
    \text{where}&~ Q' = \{(s[\vec{v} \mapsto \avec], s'[\vec{v}' \mapsto
      f(\avec)]) \mid (s,s') \in Q, \avec \in B \}
  \end{align*}
  By assumption (Formula~\ref{form:cond}), there is a coupling
  \[
    s(\dexpr) \aspace{\Psi_f} s'(\dexpr')
  \]
  so there exists a joint distribution $\mu_d \in \dist(B \times B')$ such that
  $\pi_1(\mu) = s(\dexpr)$, $\pi_2(\mu) = s'(\dexpr')$, and $\supp(\mu)
  \subseteq \Psi_f$. Thus, there is a coupling
  \[
    \sem{\vec{v} \sim \dexpr}(s) \aspace{Q'} \sem{\vec{v}' \sim \dexpr'}(s') .
  \]
  Let $\mu \in \dist(S \times S)$ be the witness of this coupling.  By
  induction, the strongest post-condition operator computes the semantics of
  deterministic commands: for any deterministic command $P$ and initial state
  $s$, $\sem{P}(s)$ is the distribution that places probability $1$ at $\post(P,
  s)$; we will abuse notation and write $\sem{P}(s)$ for the output state when
  $P$ is deterministic. Define a new witness by
  \[
    \mu' = D((s, s') \mapsto (\sem{P_D}(s), \sem{P'_D}(s')))(\mu)
  \]
  where $D$ lifts a map $S \times S \to S \times S$ to a map $\dist(S \times S)
  \to \dist(S \times S)$. Evidently,
  \[
    \supp(\mu') \subseteq \cpost(P, P', (s, s'), f)
  \]
  by the support of $\mu$ and the fact that $\post(P, -) = \sem{P_D}$,
  $\post(P_D', -) = \sem{P'_D}$. The marginal conditions $\pi_1(\mu') =
  \sem{P}(s)$ and $\pi_2(\mu') = \sem{P'}(s')$ follow from routine calculations,
  for instance
  \begin{align*}
    \pi_1(\mu')(a) &= \sum_{a'} \mu'(a, a') \\
                   &= \mu(\{ (t, t') \in S \times S \mid \sem{P_D}(t) = a \}) \\
                   &= \pi_1(\mu)(\sem{P_D}^{-1}(a)) \\
                   &= \sem{\vec{v} \sim \dexpr}(s)(\sem{P_D}^{-1}(a)) \\
                   &= \sem{P}(s)(a) .
  \end{align*}
  Therefore $\mu'$ witnesses the desired coupling
  \[
    \sem{P}(s) \aspace{\Psi} \sem{P'}(s')
  \]
  for $\Psi = \cpost(P, P', (s,s'), f)$.

\paragraph{Proof of Theorem~\ref{thm:uniform}.}
  For every pair $(\aelem, \aelem')$ of possible values for $v^*$, let
  \[
    \Psi_{\aelem, \aelem'} = \{ (s', s_1') \in S \times S \mid s'(v^*) = \aelem \iff s_1'(v^*_1) = \aelem' \} .
  \]
  Let $s$ be the input memory, and let $s_1$ be the relabeled version. We have
  \[
    \cpost(P, P_1, (s, s_1), f)
    \subseteq \cpost(P, P_1, Q, f)
    \subseteq \Psi_{\aelem, \aelem'}
  \]
  and so by Theorem~\ref{thm:cpost}, we have
  \[
    \sem{P}(s) \aspace{\Psi_{\aelem, \aelem'}} \sem{P_1}(s_1) .
  \]
  Since $P$ and $P_1$ generate the same distribution, this coupling implies that
  $v^*$ is distributed uniformly in the output distribution $\sem{P}(s)$
  (Proposition 9 from Barthe et al.~\cite{barthe2017proving}).

\paragraph{Proof of Theorem~\ref{thm:independence}.}
  For every pair $(\aelem, \aelem')$ of possible values for $v^*$, let
  \[
    \Psi_{\aelem, \aelem'}
    = \{ (s', s_1' \oplus s_2') \mid s'(w^*) = \aelem \land s'(v^*) = \aelem' \iff s_1'(w^*_1) = \aelem \land s_2'(v^*_2) = \aelem' \} .
  \]
  Let $s$ be the input memory, and let $s_1, s_2$ be two relabeled versions. We
  have
  \[
    \cpost(P, P_1; P_2, (s, s_1 \oplus s_2), f)
    \subseteq \cpost(P, P_1; P_2, Q, f)
    \subseteq \Psi_{\aelem, \aelem'}
  \]
  and so by Theorem~\ref{thm:cpost}, we have
  \[
    \sem{P}(s) \aspace{\Psi_{\aelem, \aelem'}} \sem{P_1; P_2}(s_1 \oplus s_2) .
  \]
  Since $P_1; P_2$ is the self-composition of $P$, this coupling implies that
  $w^*, v^*$ are probabilistically independent in the output distribution
  $\sem{P}(s)$ (Proposition 13 from Barthe et al.~\cite{barthe2017proving}).

\paragraph{Proof of Theorem~\ref{thm:unif-sound}.}
  Let $s_0, s_0' \in \states$ be any two states such that $s_0(v) = s_0'(v)$ for
  all $v \in V^I$, and let $b, b' \in \aset$ be any two possible values of
  $v^*$. Since $\varphi$ is satisfiable, there must exist a function $f$ such
  that \rone for all value vectors $\vec{b}, \vec{b'}$ satisfying $\vec{b'} =
  f(\vec{b})$ and output assignments $(s, s')$ to $(V, V_1)$ satisfying
  $\enc(P)$ and $\enc(P_1)$, we have $s(v^*) = b \iff s'(v^*) = b'$
  (Equation~\ref{c:main}), and \rtwo there is an $f$-coupling between
  $s_0(\dexpr) \aspace{f} s_0'(\dexpr_1)$ (Equation~\ref{c:couple}). Since by
  Lemma~\ref{lem:enc-sound} the logical encodings $\enc(P)$ and $\enc(P')$ model
  the strongest postconditions $\post(P, s_0)$ and $\post(P', s_0')$, combining
  these facts implies that
  \[
    \cpost(P, P_1, (s_0, s_0'), f)
    \subseteq
    \{ (s, s') \mid s(v^*) = b \iff s'(v^*) = b' \} .
  \]
  Since this holds for any initial states $(s_0, s_0')$ agreeing on input
  variables, we can conclude by Theorem~\ref{thm:uniform}.

\paragraph{Proof of Theorem~\ref{thm:indep-sound}.}
Similar to Theorem~\ref{thm:unif-sound}, appealing to Theorem~\ref{thm:independence}.

\paragraph{Proof of Theorem~\ref{thm:trans}.}
We first show that we can transform
a constraint of the form $\psi \triangleq \forall x \ldotp \exists f \ldotp \varphi$
into an equivalent one of the form $\psi' \triangleq \exists g \ldotp \forall x \ldotp  \varphi'$
Specifically, for each occurrence of $f(y)$ in $\varphi$,
we replace it with $g(x,y)$, where $g$ is a new uninterpreted function,
to result in $\varphi'$.

Suppose $\psi$ is valid, then for every value  $c$ of $x$,
there is an  interpretation of $f$, call it $f_c$, that
satisfies $\varphi$.
Combine all those functions into a function $g$ as follows:
\[
g(x,y) = f_c(y) \text{ if } x = c
\]
By construction, $g(x,y)$ satisfies $\forall x \ldotp  \varphi'$

Suppose $\psi'$ is valid. Then,  there is an interpretation
of $g$, call it $g'$ that satisfies the formula $\forall x \ldotp  \varphi'$.
This means that for every value $c$ of $x$,
the function $g'(c,-)$ satisfies $\varphi'$, where $g'(c,-)$ is a partial
application of $g'$ to its first argument.
It follows that $\psi$ is also valid, as
we can use $g'(c,-)$ as an interpretation of $f$ for every value $c$ of $x$.

Now consider a constraint $\dexpr \aspace{f} \dexpr'$.
Let the encoding of the constraint be $\varphi$.
Suppose $m$ is a model of $\varphi$.
By construction, $m$ sets $f$ to an injective function
such that $\forall y \ldotp m(\dexpr)(y) \leq m(\dexpr)(f(y))$.
Therefore,
$m(\dexpr) \aspace{m(f)} m(\dexpr')$ holds.

\paragraph{Proof of Theorem~\ref{thm:condindep}.}
Essentially the same as Theorems \ref{thm:independence} and
\ref{thm:unif-sound}, appealing to Proposition 16 from Barthe et
al.~\cite{barthe2017proving}.

\section{Grammar for the Search Space}\label{app:grammar}
We now discuss the search space for implementations of $f$.
By definition, $f$ is a function from $n$-tuples to $m$-tuples.
For simplicity, we assume $n = m$, which can be achieved by adding dummy (unused)
 variables in one of the programs.
 Now, our search begins by proposing the function:
 $f(x_1, \ldots, x_n) = (x_1,\ldots,x_n)$, i.e., the identity function.
 It then continues modifying it using the following set of rules, in Figure~\ref{fig:grammar},
 which populates all functions in a set $W$.
 The set $\emph{Conds}$ is the set of all Boolean conditions comprised
 of predicates appearing in the program;
 the set $\emph{Consts}$ is the set of all constants (including logical ones)
 appearing in the program.

\textsc{base} is the identity function.
\textsc{swap} swaps two of the elements in the output vector.
\textsc{neg} negates one of the output elements, assuming its type is Boolean.
\textsc{cond} generates a conditional by composing two functions.
\textsc{const} creates a constant function (this is not injective by definition,
but can be composed using other rules to create an injective function).
Our implementation enumerates programs in $W$ in a breadth-first manner,
i.e., by order of size.

\begin{figure}
  \centering
 \begin{prooftree}
 ~
 \justifies
 f(x_1,\ldots,x_n) = (x_1,\ldots,x_n) \in W
 \using \textsc{base}
 \end{prooftree}

\vspace{.3in}
 \begin{prooftree}
   f(x_1,\ldots,x_n) = (y_1,\ldots,y_n) \in W
   \quad\quad i,j \in [1,n]\quad\quad i < j
 \justifies
 f(x_1,\ldots,x_n) = (y_1,\ldots,y_{i-1},y_j,y_{i+1}, \ldots, y_{j-1},y_i,y_{j+1}, \ldots, y_n) \in W
 \using \textsc{swap}
 \end{prooftree}

 \vspace{.3in}
 \begin{prooftree}
   f(x_1,\ldots,x_n) = (y_1,\ldots,y_n) \in W
\quad\quad i \in [1,n]
 \justifies
  f(x_1,\ldots,x_n) = (y_1,\ldots,\neg y_i, \ldots,y_n) \in W
 \using \textsc{neg}
 \end{prooftree}

\vspace{.3in}
\begin{prooftree}
  f_1(x_1,\ldots,x_n) = \ldots \in W
\quad\quad
  f_2(x_1,\ldots,x_n) = \ldots \in W
\quad\quad
  C \in \emph{Conds}
\justifies
  f(x_1,\ldots,x_n) = \emph{ite}(C, f_1(x_1,\ldots,x_n), f_2(x_1,\ldots,x_n))
\using \textsc{cond}
\end{prooftree}

\vspace{.3in}
\begin{prooftree}
  c_1,\ldots,c_n \in \emph{Const}
\justifies
  f(x_1,\ldots,x_n) = (c_1,\ldots,c_n)
\using \textsc{const}
\end{prooftree}
\caption{Grammar of interpretations of $f$}\label{fig:grammar}
\end{figure}

\section{Example Encoding}\label{app:encex}
Recall the example from Section~\ref{ssec:fair}.
The encoding of the proof of uniformity of \textsf{fairCoin}
is as follows, where $V = \{x,y,p\}$ and $V_1 = \{x_1,y_1,p_1\}$:

\begin{align*}
  \forall a, a' \ldotp & \exists f, I \ldotp \forall V, V_1 \ldotp & \\
  &  p=p_1 \Longrightarrow I(\vars,\vars_1) \\
  &   I(\vars,\vars_1) \land x=y \land
   (x_1',y_1') = f(x',y') \land p' = p_1' \Longrightarrow  I(\vars',\vars_1') \\
   & I(\vars,\vars_1) \Longrightarrow (x=y \iff x_1 = y_1)\\
  %
  & I(V,V_1) \Longrightarrow \bern(p) \aspace{f} \bern(p_1) \\
  & I(V,V_1) \land  x\neq y \Longrightarrow (x = a \iff x_1 = a')
\end{align*}

\fi

\end{document}